\documentclass[useAMS,usenatbib]{mn2e}

\usepackage{graphicx,rotating,lscape,color}

\title[Strange pulsation modes in luminous red giants]
{Strange pulsation modes in luminous red giants}

\author[P. R. Wood and E. A. Olivier]{P. R. Wood$^{1}$\thanks{E-mail:
peter.wood@anu.edu.au (PRW); enrolics@gmail.com (EAO)} and
E. A. Olivier$^{2}$\\
$^{1}$Research School of Astronomy and Astrophysics, Australian National University,
Cotter Road, Weston Creek ACT 2611, Australia\\
$^{2}$Physics Department, University of Western Cape, Private bag X17, Cape Town 7535, South Africa;\\
South African Astronomicaly Observatory, PO Box 9, Observatory 7935, Cape Town, South Africa
}
\begin{document}

\date{}

\pagerange{\pageref{firstpage}--\pageref{lastpage}} \pubyear{2002}

\maketitle

\label{firstpage}

\begin{abstract}

We show that the spectrum of radial pulsation modes in luminous red
giants consists of both normal modes and a second set of modes with
periods similar to those of the normal modes.  These additional modes
are the red giant analogues of the strange modes found in classical
Cepheids and RR Lyrae variables.  Here, we describe the behaviour of
strange and normal modes in luminous red giants and discuss the
dependence of both the strange and normal modes on the outer boundary
conditions.  The strange modes always appear to be damped, 
much more so than the normal modes.  They should never be observed as
self-excited modes in real red giants but they may be detected in the
spectrum of solar-like oscillations.  A strange mode with a period
close to that of a normal mode can influence both the period and
growth rate of the normal mode.

\end{abstract}



\section{Introduction}

Luminous red giant stars are known to exhibit periods of variation that
fall on 7 or more roughly parallel period-luminosity sequences 
\citep[e.g.][]{woo99,ita04,fra05,sos07}.
Some of these sequences are known to be due to different radial pulsation modes
\citep{woo99,sos07,tak13}.  While exploring the periods and stability of radial
pulsation modes in luminous red giants, we encountered situations where we found 
two radial modes with identical pulsation periods.  It turned out that there are
two independent sets of radial pulsation modes occurring in luminous red giants,
one set being the well-known normal modes and the second set being the
strange modes encountered in classical Cepheid and RR Lyrae variables \citep*{buc97,buc01}.
Related strange modes also appear in luminous main-sequence stars \citep*{sai98}.
Here we describe the behaviour of the strange modes in red giants and how they
depend on the surface boundary conditions.

\section{The model calculations}

The study of radial pulsation in red giants requires construction of a static
model and then an analysis of the linear nonadiabatic modes of
pulsation.  A pair of computer codes is required for these two steps.
The codes used are based on those described by \citet{fox82}
and modified to include turbulent viscosity as described in
\citet{kel06} (although a turbulent viscosity parameter
$\alpha_{\rm \nu} = 0.0$ was used in the present calculations).  
Convective energy transport is treated using mixing
length theory: these models do not include turbulent pressure or the
kinetic energy of turbulent motions.  Opacities in the interior are
from the OPAL project \citep{igl96} while in the outer layers we use
the opacities of \citet{mar09} which include a molecular
component. The models use a composition X=0.73 and Z=0.008, which is
appropriate for young to intermediate age stars in the Large Magellanic Cloud (LMC).  A mixing
length of 1.97 pressure scale heights was used (to reproduce the giant
branch temperature given by \citealt{kam10} for the luminous O-rich
stars in the populous intermediate age LMC cluster NGC\,1978).

A problem with the study of red giants is that, unlike classical
pulsating stars such as Cepheids and RR Lyraes, the atmosphere is not
thin and selecting the position for the outer boundary is not
straightforward.
The outer boundary in the static models is placed according to two
requirements.  The mass zoning in the scheme used by \citet{fox82} has
radius $r$ and luminosity $L_r$ defined at zone boundaries $j$=1,...,N+1
where j=N+1 corresponds to the surface of the star.  The gas pressure
$P_{gas}$
and temperature $T$ are defined at zone centres $j+\frac{1}{2}$,
$j$=1,...,N.  \citet{fox82} use a boundary condition $P_{\rm gas}$=0 at
$j$=N+1.  In this study, we do not set the gas pressure to zero at
the surface.  Instead, our first requirement at the surface is that
$P_{\rm gas}$($j$=N+1)=0.9$P_{\rm gas}$($j$=N+$\frac{1}{2}$) so that
there is a significant gas pressure at the surface.  The factor 0.9 is 
arbitrary and is chosen so that the change in gas pressure
across the surface zone is not too large.  The second
requirement is that the optical depth from the surface ($j$=N+1) to
the centre of the outermost zone ($j$=N+$\frac{1}{2}$) is a
pre-specified value $\tau_{\rm c}$.  Given these boundary conditions and an
initial guess at $R=r$($j$=N+1), the equations of hydrostatic
structure are integrated in from the surface to the core, defined to
be at $r = 0.15\,{\rm R}_{\odot}$.  The mass of the core is defined by
this inward integration.  The outer radius $R$ is then found by iterating
on $R$ until the core mass has the required value $M_{\rm core}$.  For
AGB stars, we obtain $M_{\rm core}$ from the core mass-luminosity
relation of \citet{woo81} while for RGB stars $M_{\rm core}$ comes
from a fit to the core mass-luminosity relation of the evolutionary
models of \citet{ber08} for Z=0.008.  At this stage, the radius and
effective temperature of the model and the mass coordinates of each zone are defined:
they are dependent on the input stellar mass ($M$), luminosity ($L$)
and composition (as well as the input physics and mixing length
parameters).

\subsection{The mechanical outer boundary condition in pulsation models}

The gas pressure at the surface ($M_{\rm r}=M$) of our models is not
zero so we need to account for its variation as the defined stellar surface
oscillates.  Since we are typically studying up to 8 modes in each
star, the frequencies of the higher overtones can approach or exceed
the acoustic cutoff frequency at the surface.  This means that we need
to account for the possibility of running waves escaping through
$M_{\rm r}=M$ into the layers above.

In this study, we follow a slightly modified version of the approach
described in the Appendix, part $b$, of \citet{bak65}.  We let the
radius and pressure variations above $r = R$ be given by
$r=r_{0}(1+xe^{i \omega t})$ and $P_{\rm gas}=P_{\rm gas,0}(1+pe^{i \omega t})$, 
respectively, where the subscript 0 indicates the
static value.  We adopt the assumptions of \citet{bak65} that the
region above the surface ($M_{\rm r}=M$, $r = R$) of the star which
influences the interior pulsation is relatively small in radius compared to $R$
and it is effectively isothermal so that the gas pressure scale height
$H_0$ and the ratio $H_0/r \approx H_0/R$ are constant.  
With these approximations, the variation of $x$ with radius is given by $x
\propto e^{\nu r_{0}/R}$ where
\begin{equation}
\nu = \frac{1}{2h} \left\{ (1-4h)-\left[ (1-4h)^2-4h( \frac{4+3\sigma^2}{\gamma}-3 ) \right]^{\frac{1}{2}} \right\} .
\end{equation}
Here, $h = H_0/R$ and $\sigma^2 = \omega_{\rm r}^2/(3GM/R^3)$, where
$\omega_{\rm r}$ is the real part of $\omega$.  Without making any
further assumptions about $\nu$ (\citealt{bak65} assumed $h \ll 1$),
the general relation between $p$ and $x$ at $r=R$ is 
\begin{equation}
p = -\gamma (\nu+3)x~~.
\end{equation}
This is the mechanical boundary condition we use in our calculations.
We adopted $\gamma = 1$, corresponding to isothermal oscillations in
the outer layers.  If the expression in square brackets is negative,
then $\nu$ is complex and it corresponds to a running wave in the
region above $r = R$.  Setting the expression in square brackets to zero
defines $\sigma$ to be the acoustic cutoff frequency $\sigma_{\rm ac}$
at the adopted outer boundary of the star.  Note that unlike most
stars which have sharp boundaries with $h \ll 1$, in extended red
giants the outer boundary can be reasonably placed over a range of
radii corresponding to moderately different $R$ values.  Hence, in a
given star, $\sigma_{\rm ac}$ can vary moderately depending on where
the outer boundary is placed (see Figure~\ref{nonad+ad_p-logl} for the
value of the acoustic cutoff frequency relative to the frequency of
radial pulsation modes in a typical case).

The use of the boundary condition defined by Equations~1 and 2 gives a
smooth transition from frequencies well below the acoustic cutoff
frequency where $\nu$ is real (the condition usually assumed for
pulsating stars) to frequencies above the acoustic cutoff frequency
where $\nu$ is complex.  Note that the value of $\nu$ we use is
$\nu_{-}$ in the nomenclature of \citet{bak65}.  This value of $\nu$
gives a finite pulsation amplitude at large distances above the
stellar surface, and it corresponds to an outward propagating wave for
frequencies above the acoustic cutoff frequency.

\section{Results}

\subsection{Normal and strange modes}

\begin{figure}
\centering
\includegraphics[width=1.0\columnwidth]{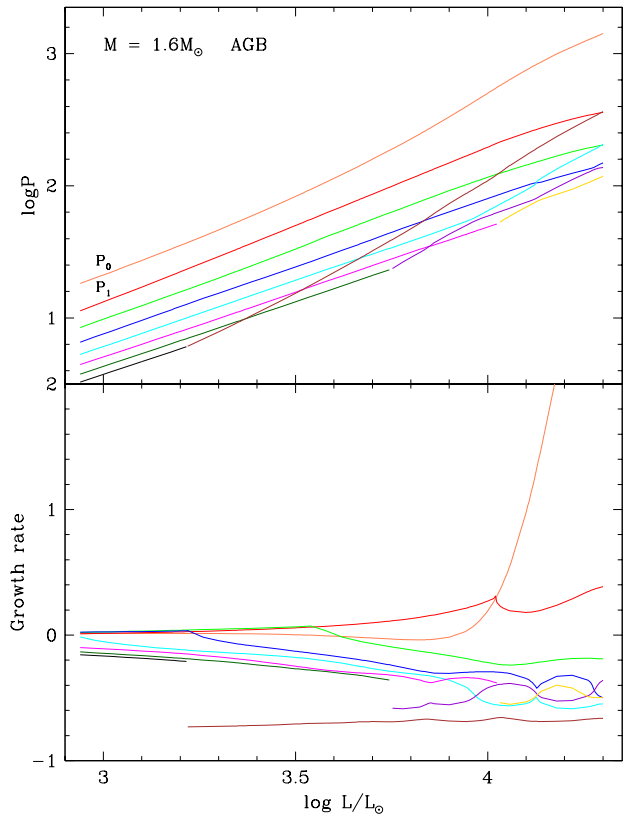}
\caption{
The periods (top panel) and growth rates (bottom panel) for the first 8 radial modes in
a series of AGB models with $M = 1.6\,{\rm M}_{\odot}$ plotted against
luminosity.  The growth rate is defined as $exp(-\omega_{\rm i}P)-1$ and it is 
the fractional growth in amplitude per period ($\omega_{\rm i}$ is the
imaginary part of the complex eigenvalue $\omega$).  
A given mode is defined by its colour with the modal colour being the
same in both panels.  See the online version for the colours.
}
\label{nonad_p+gr-logl}
\end{figure}

Figure~\ref{nonad_p+gr-logl} shows the period and growth rate for the
first 8 radial pulsation modes in a star with $M = 1.6\,{\rm
  M}_{\odot}$ as the luminosity is varied.  The optical depth to the
centre of the outer zone is set to $\log \tau_{\rm c} = -3$.  The
maximum luminosity examined for these models with 
metallicity Z=0.008 is $\log L/{\rm L}_{\odot} =
4.3$.  This is somewhat larger than the maximum observed luminosity of
$\log L/{\rm L}_{\odot} \approx 4$ for stars with $M \approx 1.6\,{\rm
  M}_{\odot}$ and metallicity ${\rm Z} \approx 0.008$ in the the
Magellanic Clouds \citep{kam10}.

At the lowest luminosities ($\log L/L_{\odot} < 3.2$), the periods and
growth rates of all modes behave smoothly as the luminosity changes.
These modes correspond to the normal modes of radial pulsation.  We
refer to the normal fundamental mode as $P_0$, the normal first
overtone as $P_1$, the normal second overtone as $P_2$ and so on.  As
the luminosity increases past $\log L/{\rm L}_{\odot} \approx 3.2$, a
mode appears with a period equal to that of $P_7$.  The damping rate
of this new mode is much higher than that of the normal mode so that
the complex eigenvalues $\omega$ of the two modes are very different.
As the luminosity of the star increases, the period of the new mode
increases more rapidly than that of the normal modes so that the
period of the new mode successively equals that of $P_6$, $P_5$, $P_4$,
$P_3$, $P_2$ and $P_1$.  However, in each case of period equality, the
growth rates and hence complex eigenvalues of the two modes are
different.

\begin{figure}
\centering
\includegraphics[width=1.0\columnwidth]{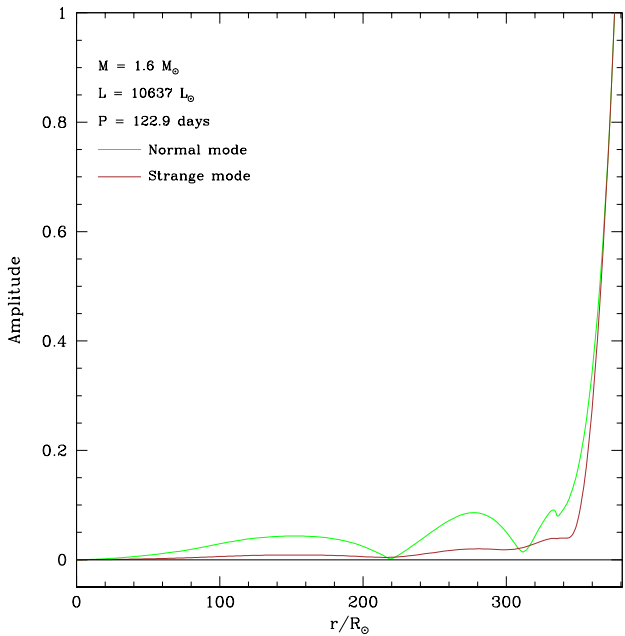}
\caption{
The amplitude of the radius perturbation eigenfunctions for a normal and strange mode of
identical period plotted against the radius $r$.  The amplitude is $(\delta_{\rm r}^2 +
\delta_{\rm i}^2)^{\frac{1}{2}}$ where $\delta_{\rm r}$ and $\delta_{\rm i}$ are the real
and imaginary parts of radius perturbation $\delta$, respectively.   
For nonadiabatic pulsation, the amplitude is not necessarily zero
at a node.  The bump in the curves at $r = 335\,{\rm R}_{\odot}$ is caused by the
hydrogen ionization zone.  
}
\label{eigenfns}
\end{figure}

The new mode belongs to the
group of modes known as strange modes.  Their behaviour is explained
lucidly in the paper by \citet{buc97} who show that the strange modes
are essentially surface modes with low interior amplitudes.
Figure~\ref{eigenfns} shows the amplitude as a function of radius for
a normal and strange mode of identical period (these two modes 
have $\log L/{\rm L}_{\odot} = 4.03$ and they lie
at the position where the brown and green lines cross in
Figure~\ref{nonad_p+gr-logl}).  It can be seen that the strange mode
is indeed more concentrated to the stellar surface than the normal
mode.  In their analysis, \citet{buc97} claimed that the periods of
normal modes and a strange modes always avoided crossing but we see no
need for this since the eigenvalues of both modes move continuously around
the complex $\omega$ plane as separate quantities.

However, avoided crossings do occur.  Following $P_4$ (the cyan line)
from low luminosities in Figure~\ref{nonad_p+gr-logl}, we see that
near $\log L/{\rm L}_{\odot} = 3.97$ the normal mode $P_4$ comes close
in period to a second strange mode (purple line).  In this case, the
two lines avoid crossing in period but they do cross in growth rate.
The effect of this is to convert the normal mode into a strange mode
and {\it vica versa}.  In fact, as the luminosity increases further,
the mode shown by the purple line undergoes another avoided crossing
and converts back from normal mode characteristics to strange mode
characteristics.  At this time, this strange mode is the third strange
mode at these luminosities (where we order the strange modes by
decreasing period).  

There are also intermediate cases where strange and normal mode
periods do cross while at the same time the growth rates are
influenced by the other mode. The growth rates tend to be attracted to
each other by mode interaction.  Examples of this behaviour are the
pink and purple modes that cross in period at $\log L/{\rm L}_{\odot}
\approx 3.85$ and the blue and cyan modes that cross in period at
$\log L/{\rm L}_{\odot} \approx 4.12$.  We note that at the high
luminosity end of the sequence of $1.6\,{\rm M}_{\odot}$ models, there
seems to be a strict alternation between normal and strange modes as
one moves to shorter periods.  This behaviour does not seem to apply
strictly at lower luminosities.

We have found no cases where strange modes have positive growth rates.
In fact, Figure~\ref{nonad_p+gr-logl} shows that the strange modes
are always more highly damped than the normal modes.
Thus, when considering self-excited modes, we should only expect to see
the normal modes of oscillation in real stars.  The strange modes
may, however, influence both the growth rate and period of
normal modes in the case of near-resonance.

It is possible that the strange modes could be stochastically
excited by convective motions and thus become observable as part of
the solar-like oscillation spectrum which has been detected in red
giants at lower luminosity \citep[e.g.][]{bed10}.  The peak power of 
a strange mode in a solar-like oscillation power spectrum will be
determined largely by the way in which the convective perturbations
can couple to the mode in question.  In addition, the peak power will
decrease as the damping of the mode increases.

An estimation of the relative amplitudes of
stochastically excited normal and strange modes is beyond the scope
of this paper.  We note that the results in \citet{ban13} show
that modes in red giants which have periods $P > 10$ days have 
relatively large amplitudes, which suggests that these modes are self-excitated.   
On the other hand, the shorter period, lower
amplitude modes appear to be stochastically excited.  Thus in the
luminous red giants considered here, which have $P > 10$ days for at
least the first two modes, it may be difficult to find the signal of
a very damped strange mode in the overall power spectrum 
where excited modes are likely to dominate.  However, if the signal of a strange
mode could be detected in a given star, its strange mode nature
could possibly be determined by its frequency spacing relative to
nearby radial ($\ell = 0$) modes.

\begin{figure}
\centering
\includegraphics[width=1.0\columnwidth]{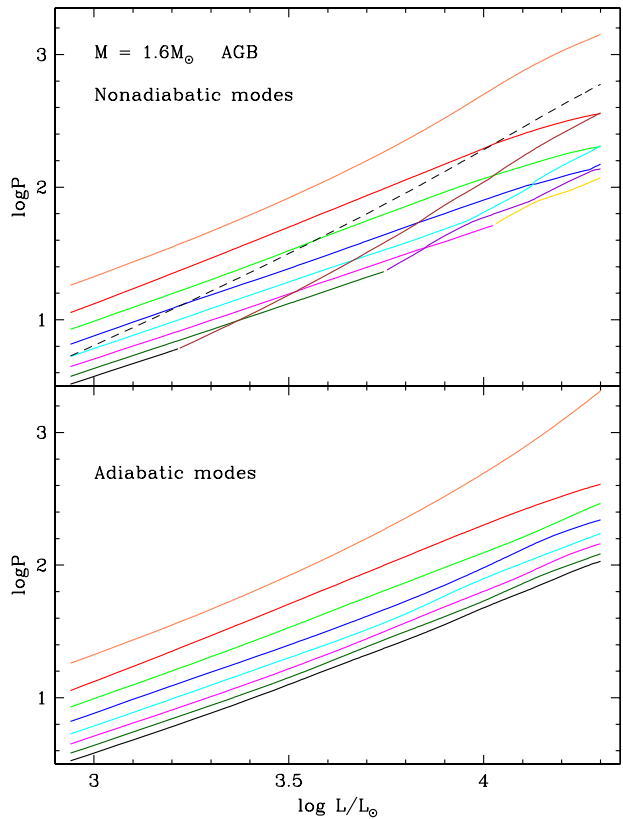}
\caption{
Top panel: Similar to the top panel in Figure~\ref{nonad_p+gr-logl} but also showing the acoustic cutoff
period corresponding to the acoustic cutoff frequency $\sigma_{\rm ac}$ at the surface of the 
star (dashed black line).  Bottom panel:  The periods of the first 8 adiabatic radial modes.
}
\label{nonad+ad_p-logl}
\end{figure}

Although strange modes and normal modes can have equal periods in the
nonadiabatic case, for adiabatic pulsation the Sturm-Liouville theorem
requires that the eigenvalues $\omega$ remain distinct.  This is shown
in Figure~\ref{nonad+ad_p-logl}.  However, strange mode behaviour
still influences the periods to some extent as seen most prominently
for $P_2$ and $P_3$ (the green and blue modes) near $\log L/{\rm
  L}_{\odot} \approx 4.15$.  A full discussion of the origin of
strange mode behaviour in the adiabatic case in given in
\citet{buc97}.

\subsection{The effect of the position of the outer boundary}
 
\begin{figure}
\centering
\includegraphics[width=1.0\columnwidth]{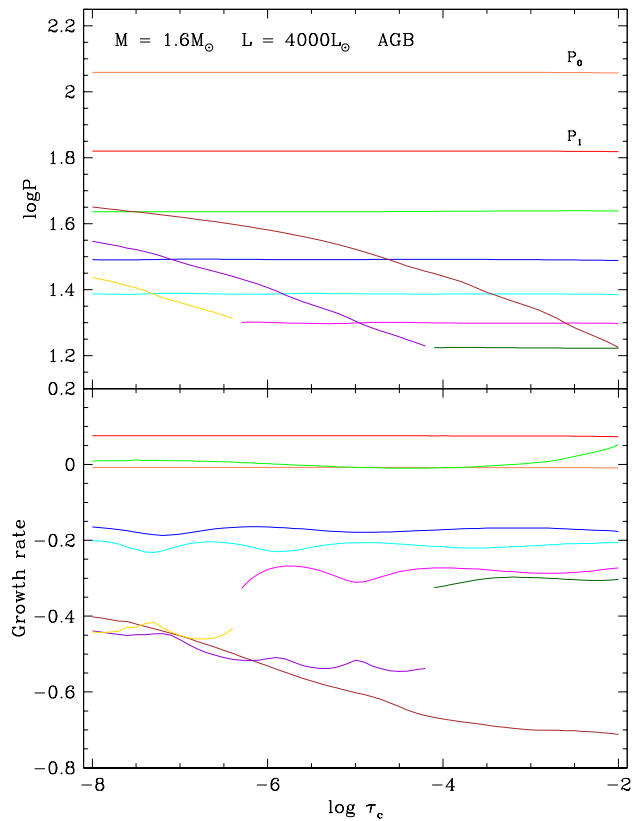}
\caption{
Similar to Figure~\ref{nonad_p+gr-logl} except that all models
have $L = 4000\,{\rm L}_{\odot}$ and the periods and growth rates are
plotted against $\log \tau_{\rm c}$ rather than $\log L/{\rm L}_{\odot}$. 
}
\label{nonad_p+gr-tau}
\end{figure}

\begin{figure}
\centering
\includegraphics[width=1.0\columnwidth]{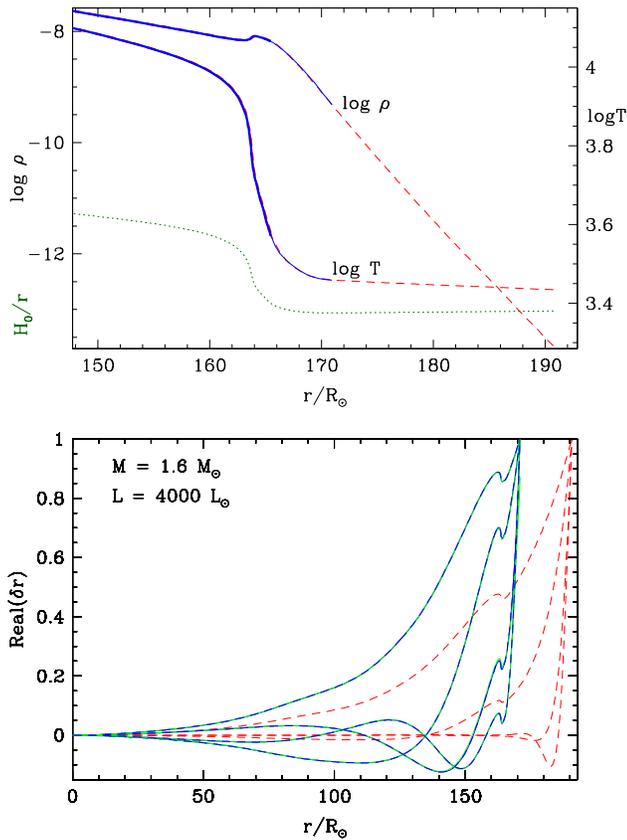}
\caption{
Surface structure and eigenfunctions for models with $M = 1.6\,{\rm
M}_{\odot}$ and $L = 4000\,{\rm L}_{\odot}$.  Top panel: $\log \rho$
and $\log T$ plotted against $r$/R$_{\odot}$ for a model with 
$\log \tau_{\rm c} = -3$ (solid blue lines) and a model with $\log \tau_{\rm c} = -8$
(dashed red lines).  The thick parts of the line correspond to
regions that are convective.  The green dotted line shows the ratio of
the pressure scale height H$_0$ to $r$, with a vertical axis scale from
0 to 1.  Bottom panel: the real part of the radius eigenfunctions for
the first 4 modes of pulsation plotted against the radius $r$.  Blue
solid lines show the eigenfunctions for the model with $\log \tau_{\rm c} = -3$
while the red dashed lines show the eigenfunctions for the model
with $\log \tau_{\rm c} = -8$.  These complex eigenfunctions are normalized
to 1.0 at the surface for both models.  The green dashed lines show
the eigenfunctions for the model with $\log \tau_{\rm c} = -8$ normalized to
1.0 at a radius which corresponds to the surface of the model with 
$\log \tau_{\rm c} = -3$.
}
\label{eigfn}
\end{figure}

We now show how the placement of the outer boundary influences the
pulsation periods of red giants.  
The periods of the eight lowest order modes in a star with $M = 1.6\,{\rm
  M}_{\odot}$ and $L = 4000\,{\rm L}_{\odot}$ are plotted against
$\log \tau_{\rm c}$ in the Figure~\ref{nonad_p+gr-tau}.  Note that a
decrease in $\tau_{\rm c}$ means that the surface radius is placed
further out in the stellar atmosphere (see below).  We
restricted $\log \tau_{\rm c} < -2$ since for larger values of
$\tau_{\rm c}$ the outer boundary is placed in a region where the
temperature gradient starts to become significant and the periods of
all modes become dependent of $\tau_{\rm c}$.

It is clear that the periods of the normal modes are independent of
$\log \tau_{\rm c}$ i.e. they are independent of the placement of the
stellar surface.  For the strange modes, the periods vary markedly
with $\log \tau_{\rm c}$.  This is consistent with the fact that the
strange modes are predominantly surface modes largely confined to the
region between the hydrogen ionization zone and the stellar surface
\citep{buc97,sai98}.  The reason that the periods of the strange
modes increase as the outer boundary is placed at larger radii
is that the ratio $z_0$ of the radius of the hydrogen ionization zone
to the stellar surface decreases.  As shown in the toy models
of \citet{buc97} (see their Figure 13), decreasing $z_0$ causes the strange mode period
to increase relative to the normal mode periods.  Note also that it is the decrease
in $z_0$ with luminosity in the sequence of 1.6\,M$_{\odot}$ models
shown in Figure~\ref{nonad_p+gr-logl} that causes the strange
mode periods to increase faster with luminosity than the normal mode
periods.

The strange modes in Figure~\ref{nonad_p+gr-tau} all
have lower growth rates (larger damping rates) than the normal modes.
They also cross the periods of the normal modes and in each case it
can be seen that mode interaction influences the growth rate as the
modes come in and out of resonance.

The top panel of Figure~\ref{eigfn}
shows the outer structure of a typical luminous red giant 
from the centre to the surface for two placements of
the outer boundary, corresponding to $\log \tau_{\rm c} = -3$ and
$\log \tau_{\rm c} = -8$.  The model with $\log \tau_{\rm c} = -8$ has
a considerably larger surface radius than the model with $\log
\tau_{\rm c} = -3$.  The physical structure of the two models is
essentially indistinguishable at common radii.

The eigenfunctions of the radius perturbation for the 4 lowest order
modes are shown in the bottom panel of Figure~\ref{eigfn} for
the two placements of the outer boundary.  At first
sight, the eigenfunctions of the corresponding modes in the two models
with different surface radii look very different.  However, this is not so.  To
compare the eigenfunctions with the same normalization, the
eigenfunctions for the more extended model were multiplied by a complex
constant which caused the eigenfunction to be normalized to a value of 1.0 at a
radius corresponding to the surface radius of the smaller model.  These
transformed eigenfunctions are shown as green dashed lines in
Figure~\ref{eigfn}.  It can be seen that these transformed
eigenfunctions are essentially indistinguishable from those of the 
less extended model (blue lines) at common radii.  

It is not clear to us where the outer boundary should be placed.
Clearly, it should be at $\log \tau_{\rm c} < -2$ since the periods of
all modes are affected if the outermost model point is deeper in the
star than the surface point of the model with $\log \tau_{\rm c} =
-2$.  
Since the periods of the normal modes are essentially
independent of the position adopted for the stellar surface, we have
adopted $\log \tau_{\rm c} = -3$ as the criterion by which our outer
boundary is defined.  This value of $\tau_{\rm c}$ also means that the
growth rates of the normal modes are not greatly affected by mode
interaction with strange modes.  As seen in
Figure~\ref{nonad_p+gr-tau}, these interactions can become significant
for very small $\tau_{\rm c}$ values when the eigenfunctions extend
high into the stellar atmosphere.

The strong dependence of strange mode periods on the adopted
surface radius means that if strange modes could be detected
in the spectrum of solar-like oscillations in a red giant, then the
detected periods could be used to determine the outer radius
of the red giant as experienced by strange modes.

\section{Summary and conclusions}

We have shown that in luminous red giant stars, a series of radial
strange modes exists in addition to the series of radial normal modes
of pulsation.  At high luminosities, strange modes can have periods as
long as that of the first overtone for plausible luminosities,
especially if an extended outer atmosphere in included in the
calculations.  The periods of the strange modes increase faster with
$\log L/{\rm L}_{\odot}$ (and hence surface radius) than the periods
of the normal modes.  This means that normal and strange modes in a
given star can have identical periods at certain luminosities.  In
some cases, avoided crossings in period occur leading to a given mode
(identified by continuity as the luminosity is varied) changing back
and forth between a normal and strange mode character as the
luminosity changes.  In cases where the modes cross in period, the
growth rates of each mode is affected by the near-resonance condition.
The strange modes are always damped, and more so than the normal modes. 
We should not expect to see self-excited strange modes in real
stars but strange modes may be observed in the spectrum of solar-like oscillations.
The periods and growth rates of normal modes may be influened by
resonances with strange modes.  Fortunately, the normal modes periods
are essentially unaffected by the placement of the outer boundary.  On
the other hand, the strange mode periods increase as the outer
boundary is placed at larger radii.  Finally, we note that although
these calculations were performed for radial modes, we expect that
strange modes should also exist in the nonradial case.

\section*{Acknowledgments}

PRW was partially funded during this research by the Australian
Research Council Discovery grant DP1095368.  He was also gratefully
acknowledges funding from UWC and SAAO for travel and accommodation
during a trip to South Africa where some of this work was done.
We thank the anonymous referee for useful comments.


\bsp

\label{lastpage}

\end{document}